\documentclass[pra,showpacs,twocolumn]{revtex4-1}

\usepackage{amsfonts,amssymb, amsmath,graphicx,mathrsfs,bm,appendix}
\usepackage{float}
\usepackage{txfonts}
\begin{document}

\title{Nonlocal coupling effects on single-photon transmission in a one-dimensional waveguide interacted with a side optical cavity}

\author{Yongyou Zhang}
\email[Author to whom correspondence should be addressed. Electronic mail: ]{yyzhang@bit.edu.cn}
\author{Bingsuo Zou}
\affiliation{Beijing Key Lab of Nanophotonics $\&$ Ultrafine Optoelectronic Systems and School of Physics, Beijing Institute of Technology, Beijing 100081, China}%

\begin{abstract}
  The nonlocal coupling effect between a one-dimensional waveguide (1DW) and a side optical cavity (SOC) is studied. We first find the real-space Hamiltonian of the nonlocal-coupling system of the 1DW and SOC, and then derive out an equation determining the energy of the hybridization state between the 1DW and SOC modes and an analytic formula for the single-photon transmission. Through them, we recognize that the single-photon transmission-dip position can be changed by adjusting the SOC size and the coupling strength between the 1DW and SOC. The transmission spectra strongly depends on the nonlocal-coupling function between the 1DW and SOC, and holds an asymmetry line shape. At last, we simulate the asymmetry of the single-photon transmission by a gold-based waveguide coupled with a rectangular SOC.
\end{abstract}

\pacs{42.50.Ct, 42.79.Gn, 42.50.Pq}
\maketitle

\section{introduction}

In recent years, researchers have studied the photon transmission and correlation in the one-dimensional waveguide coupling to a wide
variety of quantum systems, including quantum dots \cite{cheng2012fano, Englund2010Resonant}, a cavity with an atom inside \cite{Shi2011Twophoton, Ji2012Twophoton} or Kerr nonlinearity \cite{Liao2010Correlated}, single or multiple atoms holding two or multiple levels \cite{Shen2007Strongly1, Shen2007Strongly2, Roy2011TwoPhoton, Roy2013Twophoton, Fan2010Inputoutput, Rephaeli2011Fewphoton, Longo2011Fewphoton, Kolchin2011Nonlinear, Zheng2012Strongly}, and an optomechanical cavity \cite{Liao2013Correlated}. This is due to their strong potential in manipulating and controlling the photon-state transmission \cite{Huang2013Controlling, huang2011subwavelength, Kekatpure2010Phase, chen2012multiple, han2011plasmon, dong2013group, Shen2007Stopping, shen2009theory1, shen2009theory2, tan2011entangling, chang2007single, tu2010coupled, xiao2010asymmetric, lu2012plasmonic, Pan2010Experimental, sandhu2010enhancing}. Plenty of significant experimental efforts have been done to realize these systems on chip in integrated structures \cite{ Englund2010Resonant, Birnbaum2005Photon, Lang2011Observation, Dayan2008APhoton}. In these structures, many interesting phenomena are observed, for example, electromagnetically induced transparency \cite{huang2011subwavelength, Kekatpure2010Phase, chen2012multiple,han2011plasmon}, Fano resonance \cite{xiao2010asymmetric,tu2010coupled,lu2012plasmonic}, slow light behavior \cite{huang2011subwavelength,dong2013group}, multi-photon transmission\cite{Gullans2013Single, Liao2013Correlated, Shi2013Twophoton, Liao2010Correlated, Xu2013Analytic}, and so on \cite{Huang2013Photon}. With these phenomena, researchers have designed optical switches \cite{Fan2002Sharp, sandhu2010enhancing}, single photon transistors \cite{chang2007single, Neumeier2013Single}, photon memory \cite{huang2011subwavelength}, and band filters \cite{pan2010tuning}. They potentially accelerate the development of the optical quantum information processing.

In these previous works, researchers generally adopted the on-site coupling model described by the Dirac function to express the interaction between the 1D waveguide (1DW) and side optical cavity (SOC) \cite{shen2009theory1, shen2009theory2, Shen2007Stopping}. This model is valid when the transversal (parallel to the 1DW) scale $w$ of the SOC is much narrower than the wavelength $\lambda$ of the guided light, as shown in Fig.~1(a) where $w\ll\lambda$. On the contrary, it might not be suitable for the case given in Fig.~1(b) where the transversal scale of the SOC is comparable with or larger than the wavelength of the guide light, namely, $w\gtrsim\lambda$. This is owing to that the SOC mode cannot be assumed as the local one compared to the waveguide's. The cavity photon can tunnel into the waveguide not only through SOC center region but also through its side places, as shown in Fig.~1(b). Hence, the coupling between the 1DW and SOC for the structure shown in Fig.~1(b) could not be described by the local model, as well as the Dirac function. Instead, the coupling depending on the transversal coordinate should be used.

\begin{figure}
  \centering
  \includegraphics[width=6.5 cm]{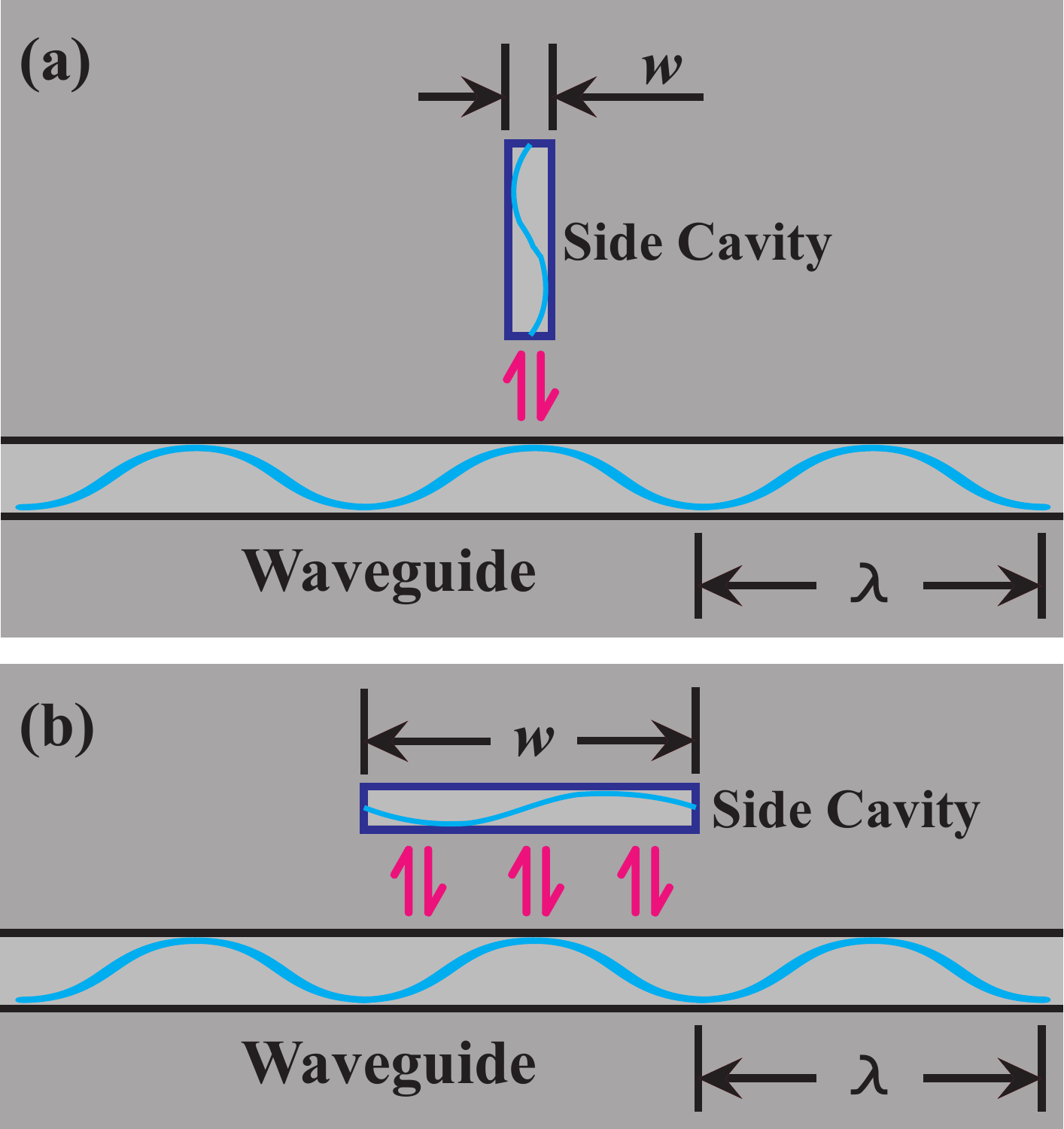}
  \caption{Schematics of a waveguide side coupled with one narrow cavity (a), and one wide cavity (b). The wave curves represent the photon modes in the cavity and the waveguide, respectively. The arrows denote the possible photon tunnel between the side cavity and the waveguide. $\lambda$ and $w$ are the waveguide photon wavelength and the width of the side cavity, respectively.}
\end{figure}

In this work, as an extension of the on-site model, we consider the single-photon transmission of the coupling structure of the 1DW and SOC between which the interaction depends on the transversal coordinate. A quantum field method is used to find the dependence of the interaction between the 1DW and SOC on the transversal coordinate. We show that it is mainly determined by the SOC mode, and that it can strongly influences the single-photon transmission and the hybridization state between the 1DW and SOC modes. Besides, we derive out an analytical expression for the single-photon transmission and a dispersion equation for the hybridization state. For convenience, the coupling structure of the 1DW and SOC is abbreviated as 1DW-SOC hereafter.

This work is organized as follows. In Sec.~II, we first find the real-space Hamiltonian for the structure 1DW-SOC from the quantum field theory, and then derive the hybridization state between the 1DW and SOC modes and the single-photon transmission formula. Then, in Sec.~III we discuss two coupling cases: zeroth-order (Gaussian) coupling, and first-order coupling. The former is described by the Gaussian function, and the second is described by the first derivative of the Gaussian function. In Sec.~IV, a gold-based 1DW-SOC structure is taken to simulate the nonlocal effects of the single-photon transmission. At last, a brief conclusion is summarized in Sec.~V.

\section{Model and Formulas}

\subsection{Nonlocal coupling Hamiltonian}
Because the considered structure shown in Fig.~1, strictly speaking, is two dimensional, we need to start from the two-dimensional optical structure to find its Hamiltonian. Please note that though our start point is the two-dimensional structure, the following developed theory is also valid for the three dimensional situations. This can be seen from the following text. For the two-dimensional optical structure, the photon Hamiltonian can be written as
\begin{align}
{\cal H}=&\iint dxdy\hat{\psi}^\dag(x,y)\hat{\omega}_{ph}(-i\nabla)\hat{\psi}(x,y)
\end{align}
with the help of the photon energy operator $\hat{\omega}_{ph}(-i\nabla)$ and the quantum photon field operator $\hat{\psi}(x,y)$. For the coupling structure of the 1DW-SOC in Fig.~1, the quantum photon field can be divided into three parts, namely, the right-moving waveguide field $\hat{\psi}_R(x,y)$, the left-moving waveguide field $\hat{\psi}_L(x,y)$, and the cavity field $\hat{\psi}_C(x,y)$, that is,
\begin{align}
\hat{\psi}(x,y)=\hat{\psi}_R(x,y)+\hat{\psi}_L(x,y)+\hat{\psi}_C(x,y).
\end{align}
For the structure given in Fig.~1, the cavity is limited in both directions $x$ and $y$, and that the waveguide is limited only in direction $y$. Therefore, we can consider only one cavity mode and one corresponding waveguide mode. As a result, three quantum fields can be cast into the following forms
\begin{subequations}
\begin{align}
\hat{\psi}_R(x,y)&=\hat{\psi}_R(x)\phi_w(y),\\
\hat{\psi}_L(x,y)&=\hat{\psi}_L(x)\phi_w(y),\\
\hat{\psi}_C(x,y)&=\hat{c}\phi_c(x,y).
\end{align}
\end{subequations}
$\hat{\psi}_{R(L)}(x)$ and $\hat{c}$ are the right-moving (left-moving) photon field operator and the cavity photon annihilation operator, respectively. $\phi_w(y)$ and $\phi_c(x,y)$ are the normalized eigenfunctions of the waveguide and cavity. They satisfy
\begin{subequations}
\begin{align}
\hat{\omega}_{ph}(-i\nabla)\phi_w(y)&=\hat{\omega}_{ph}\left(-i{\partial \over \partial x}\right)\phi_w(y),\\
\hat{\omega}_{ph}(-i\nabla)\phi_c(x,y-d)&=(\omega_c-i\gamma_c)\phi_c(x,y).
\end{align}
\end{subequations}
Here, $\omega_c$ and $\gamma_c$ are the energy and loss rate of the SOC, respectively. Taking Eqs.~(2-4) into the the Hamiltonian, we can have
\begin{align}
{\cal H}&{=}\int dx \left[\hat{\psi}_R^\dag(x)\hat{\omega}_{ph}\left(-i{\partial \over \partial x}\right)\hat{\psi}_R(x) \right.\nonumber\\ &{+}\left.\hat{\psi}_L^\dag(x)\hat{\omega}_{ph}\left(+i{\partial \over \partial x}\right) \hat{\psi}_L(x)\right] +(\omega_c{-}i\gamma_c)\hat{c}^\dag \hat{c}\nonumber\\
&{+}\!\int dx\left\{V(x)\left[\hat{\psi}_R^\dag(x){+}\hat{\psi}_L^\dag(x)\right]\hat{c} {+}V^*(x)\hat{c}^\dag\left[\hat{\psi}_R(x){+}\hat{\psi}_L(x)\right]\right\}
\end{align}
where
\begin{align}
\hat{\omega}_{ph}\left(\pm i{\partial \over \partial x}\right)=\left(\omega_c-v_gk_c\right)\pm iv_g\frac{\partial}{\partial x},\\
V(x)=(\omega_c{-}i\gamma_c)\int dy \phi_w^*(y)\phi_c(x,y).
\end{align}
For convenience, the Planck constant has been set to be $\hbar=1$ here.
The dispersion of the waveguide mode has been linearized by introducing a group velocity $v_g$ at the wave vector of $k_c$, namely, the waveguide photon energy is $\varepsilon=(\omega_c-v_gk_c) \pm v_gk$ around the energy point of $\omega_c$ for the right-moving and left-moving photons, respectively. Both right and left-moving photons have the same frequency $\omega_c$ when their wave vectors equal $\pm k_c$, respectively.
The term $V(x)$ represents the mode coupling between the 1DW and SOC. When $\phi_c(x,y)$ is even on coordinate $x$ and its distribution is narrow enough along direction $x$, the coupling $V(x)$ can be approximated by a Dirac function $\delta(x)$. This implies that the transversal scale $w$ of the SOC is much smaller than the photon wavelength $\lambda$, namely, $w\ll\lambda$. In this limitation, the Hamiltonian in Eq.~(5) will go back to the well studied one \cite{shen2009theory1, shen2009theory2}. In addition, the Hamiltonian in Eq.~(1), as well as the Eq.~(5),x can be extended into three dimensional situations conveniently by taking that the waveguide is limited in directions $y$ and $z$ and that the cavity is limited in all directions $x$, $y$, and $z$.

\subsection{Single photon input}

For the single photon input, the eigenstate of $\cal H$ takes the form of
\begin{equation}
|\Psi\rangle{=}C\hat{c}^\dag|\varnothing\rangle{+}\int_{-\infty}^\infty dx\left[\Phi _R(x)\hat{\psi}_R^\dag(x){+}\Phi_L(x)\hat{\psi}_L^\dag(x)\right]|\varnothing\rangle
\end{equation}
where $|\varnothing\rangle$ represents the vacuum state, with zero photon in the cavity and waveguide. $C$ is the excitation amplitude of the SOC. $\Phi_L(x)$ and $\Phi_R(x)$ are the single-photon wave functions of the left- and right-moving modes, respectively. Substituting Eqs.~(5) and (8) into the eigenvalue equation
\begin{equation}
{\cal H}|\Psi\rangle=\varepsilon|\Psi\rangle
\end{equation}
with $\varepsilon$ being the energy of the incident photon, we can find the coupled equations for $\Phi_R(x)$, $\Phi_L(x)$, and $C$ as follows
\begin{subequations}
\begin{align}
-iv_g{\partial\over\partial x}\Phi_R(x)+ V(x)C=\left(\varepsilon{-}\omega_c{+}v_gk_c\right)\Phi_R(x),\\
iv_g{\partial\over\partial x}\Phi_L(x)+ V(x)C=\left(\varepsilon{-}\omega_c{+}v_gk_c\right)\Phi_L(x),\\
\int_{-\infty}^\infty dx V^*(x)\left[\Phi_R(x)+\Phi_L(x)\right]=\left(\varepsilon-\omega_c+i\gamma_c\right)C.
\end{align}
\end{subequations}
Because of the delocalization of the coupling function, we do the following transformation for $\Phi_R(x)$ and $\Phi_L(x)$, namely,
\begin{subequations}
\begin{align}
\Phi_R(x)&=\xi_R(x)e^{ikx},\\
\Phi_L(x)&=\xi_L(x)e^{-ikx}.
\end{align}
\end{subequations}
Casting them into Eq.~(10), we get
\begin{subequations}
\begin{align}
&\xi_R(x)=\xi_R(-\infty)-iv_g^{-1}V_k(x)C,\\
&\xi_L(x)=\xi_L(-\infty)+iv_g^{-1}V_{-k}(x)C,\\
&C={V_k^*\xi_R(-\infty)+V_{-k}^*\xi_L(-\infty)\over \varepsilon-(\omega_c+\Delta_k)+i(\gamma_c-\gamma_k)}.
\end{align}
\end{subequations}
Here,
\begin{subequations}
\begin{align}
V_k(x)&=\int_{-\infty}^x dx^\prime V(x^\prime)e^{-ikx^\prime},\\
V_k&=V_k(\infty),\\
\Delta_k&={2\over v_g}\int_0^\infty dx\sin(kx)Q_r(x),\\
\gamma_k&={2\over v_g}\int_0^\infty dx\sin(kx)Q_i(x).
\end{align}
\end{subequations}
The functions of $Q_r(x)$ and $Q_i(x)$ are the real and imaginary parts of the function $Q(x)$ which is defined as
\begin{align}
Q(x)=\int_{-\infty}^\infty dx^\prime\left[V^*(x^\prime)V(x^\prime-x)\right]
\end{align}
and satisfies
\begin{align}
Q(-x)=Q^*(x).
\end{align}
When the coupling $V(x)$ holds certain parity, namely, $V(-x)=V(x)$ or $V(-x)=-V(x)$, we have $Q_i(x)=0$ and simultaneously $\gamma_k=0$. From Eq.~(12), we can find the superposition state of the 1DW and SOC modes and the single photon transmission with different boundary conditions for the functions of $\xi_R(x)$ and $\xi_L(x)$.

\subsection{Hybridization mode}

The superposition state of the 1DW and SOC modes can also be called as the hybridization state (HS). In the HS, the functions of $\xi_R(x)$ and $\xi_L(x)$ satisfy the following boundary conditions
\begin{subequations}
\begin{align}
\xi_R(x)|_{x\rightarrow\infty}&=\xi_R(x)|_{x\rightarrow-\infty}=0,\\
\xi_L(x)|_{x\rightarrow\infty}&=\xi_L(x)|_{x\rightarrow-\infty}=0.
\end{align}
\end{subequations}
Using Eq.~(12), one can find the energy of the nontrivial HS, $\varepsilon_h$, and the corresponding wave vector, $k_h$. They are determined by the equation set,
\begin{subequations}
\begin{align}
\varepsilon_h&=\omega_c+\Delta_{k_h}-i(\gamma_c-\gamma_{k_h}),\\
{\rm Re}[\varepsilon_h]&=(\omega_c-v_gk_c) + v_gk_h.
\end{align}
\end{subequations}
When the coupling function is proportional to the Dirac function, namely, $V(x)\propto\delta(x)$, the function $Q(x)$ will also be proportional to $\delta(x)$ [refer to Eq.~(14)], and therefore $\Delta_k=\gamma_k=0$ [refer to Eqs.~(13c) and (13d)]. As a result, the HS energy $\varepsilon_h$ just equals the one of the SOC mode, referred to Eq.~(17a). This implies that the on-site coupling (described by the Dirac function) between the 1DW and SOC does not shift the HS energy from the SOC mode. On the contrary, one can expect that the nonlocal coupling could induce the deviation of the HS energy from the SOC mode.

Because the Hamiltonian in Eq.~(5) is the one linearized near the energy of the SOC mode, it is valid only when the photon wave vector $k$ is near $k_c$. Based on this, the energy deviation $\Delta_{k_h}$ of the hybridization mode to the SOC mode can be roughly approximated by $\Delta_{k_c}$.

\subsection{Single photon transmission}

In the present subsection, our aim is to find the single photon transmissivity. The corresponding boundary conditions for the functions of $\xi_R(x)$ and $\xi_L(x)$ are
\begin{subequations}
\begin{align}
\xi_R(-\infty)=&1,&\xi_R(\infty)&=t,\\
\xi_L(-\infty)=&r,&\xi_L(\infty)&=0,
\end{align}
\end{subequations}
where $t$ and $r$ are the transmission and reflection coefficients, respectively.
With these boundary conditions, one can find $r$ and $t$ from Eq.~(12) as follows
\begin{subequations}
\begin{align}
r&=-i{v_g^{-1}V_k^*V_{-k} \over \varepsilon-(\omega_c+\Delta_k)+i(\gamma_c-\gamma_k)+iv_g^{-1}|V_k|^2},\\
t&=1+{V_k\over V_{-k}}r.
\end{align}
\end{subequations}
Referred to Eqs.~(13a) and (13b), when the SOC mode is even on coordinate $x$, one has $V_k=V_{-k}$ and thus $t=1+r$; oppositely, when the SOC mode is odd, one has $V_k=-V_{-k}$ and $t=1-r$. The on-site coupling described by the Dirac function is a special case of the previous situation \cite{shen2009theory1}, and thus holds the relation $t=1+r$.

In order to compare with the on-site coupling, namely, $V(x)\propto\delta(x)$, it is convenient to cast $t$ into the form
\begin{align}
t={1 \over 1+i{J_k\over \varepsilon-(\omega_c+\Delta_k)+i(\gamma_c-\gamma_k)}}
\end{align}
where
\begin{align}
J_k&={1\over v_g}|V_k|^2.
\end{align}
When $V(x)=V_0\delta(x)$, one can easily find
$$
J_k=V_0^2/v_g,~~~~~\Delta_k=\gamma_k=0,
$$
which are the same to the previous works \cite{shen2009theory1, dong2013group}. This indicates that the nonlocal-coupling influence on the single-photon transmission is fully included in $J_k$, $\Delta_k$ and $\gamma_k$. $J_k$ measures the coupling strength of the SOC mode with the 1DW mode whose wavevector is $k$. $\Delta_k$ induces the shift of the resonant-coupling position from the SOC mode. The shift value strongly depends on the form of $V(x)$, and thus strongly on the cavity mode, referred to Eqs.~(7), (13c) and (14).

The reflectivity $R$ and transmissivity $T$ can be directly obtained from
\begin{align}
R=|r|^2,~~~~~~~~~~~~~~~~~T=|t|^2.
\end{align}

\section{Examples of nonlocal coupling}

The nonlocal coupling $V(x)$ could induce different transmission behavior in the 1DW-SOC structure. In the present section, we take two typical coupling functions for $V(x)$ to show the nonlocal coupling effects, namely, the zeroth-order coupling (ZOC) and the first-order coupling (FOC), denoted as $V^z(x)$ and $V^f(x)$, respectively. They are defined as
\begin{subequations}
\begin{align}
V^z(x)&=V_0\times\left({1\over4}\pi w^2\right)^{-1/4}e^{-2x^2/w^2},\\
V^f(x)&=V_0\times\left({1\over4}\pi w^2\right)^{-1/4}\sqrt{8}{x\over w}e^{-2x^2/w^2}
\end{align}
\end{subequations}
where $V_0$ and $w$ represent the coupling strength and the SOC width. Note that $V(x)$ is determined by the SOC mode, referred to Eq.~(7). Hence, we normalize their squares to $V_0^2$. Besides, $V^z(x)$ and $V^f(x)$ correspond to the even and odd SOC modes, respectively. For convenience, hereafter we will take the units of energy (or frequency), wave vector, length as $\omega_c$, $k_c$, and $\lambda_c=2\pi/k_c$, respectively.

The Fourier transforms of $V^z(x)$ and $V^f(x)$ are found as follows
\begin{subequations}
\begin{align}
V_k^z&=V_0\times\left(\pi w^2\right)^{1/4}e^{-k^2w^2/8},\\
V_k^f&=V_0\times\left({1\over4}\pi w^2\right)^{1/4}\left(-ikw\right)e^{-k^2w^2/8}.
\end{align}
\end{subequations}
With the help of the coupling function, one can get the HS energy $\varepsilon_h$, nonlocal coupling $J_k$, single-photon transmission $t$ from Eqs.~(17), (20), and (21). By some algebra derivation, we can obtain that the HS wave vectors for the ZOC, $k_h^z$, and FOC, $k_h^f$, are determined by
\begin{subequations}
\begin{align}
V_0^2&={1\over2}v_g^2\left(k_h^z-k_c\right)w^{-1} \left[{\cal F}\left({1\over2} k_h^z w\right)\right]^{-1},\\
V_0^2&=v_g^2k_h^{f-1}\left(k_h^f-k_c\right)w^{-2} \left[k_h^f w{\cal F}\left({1\over2} k_h^z w\right)-1\right]^{-1},
\end{align}
\end{subequations}
respectively. Here, ${\cal F}(\alpha)$ is the Dawson integral, defined as ${\cal F}(\alpha)= \exp(-\alpha^2)\int_0^\alpha \exp(\beta^2)d\beta$. In order to find $k_h$ for the ZOC and FOC, the coupling constant $V_0$ is plotted as the function of $k_h$ and $w$ in Fig.~2 with the help of Eq.~(25).

\begin{figure}
  \centering
  \includegraphics[width=8.5 cm]{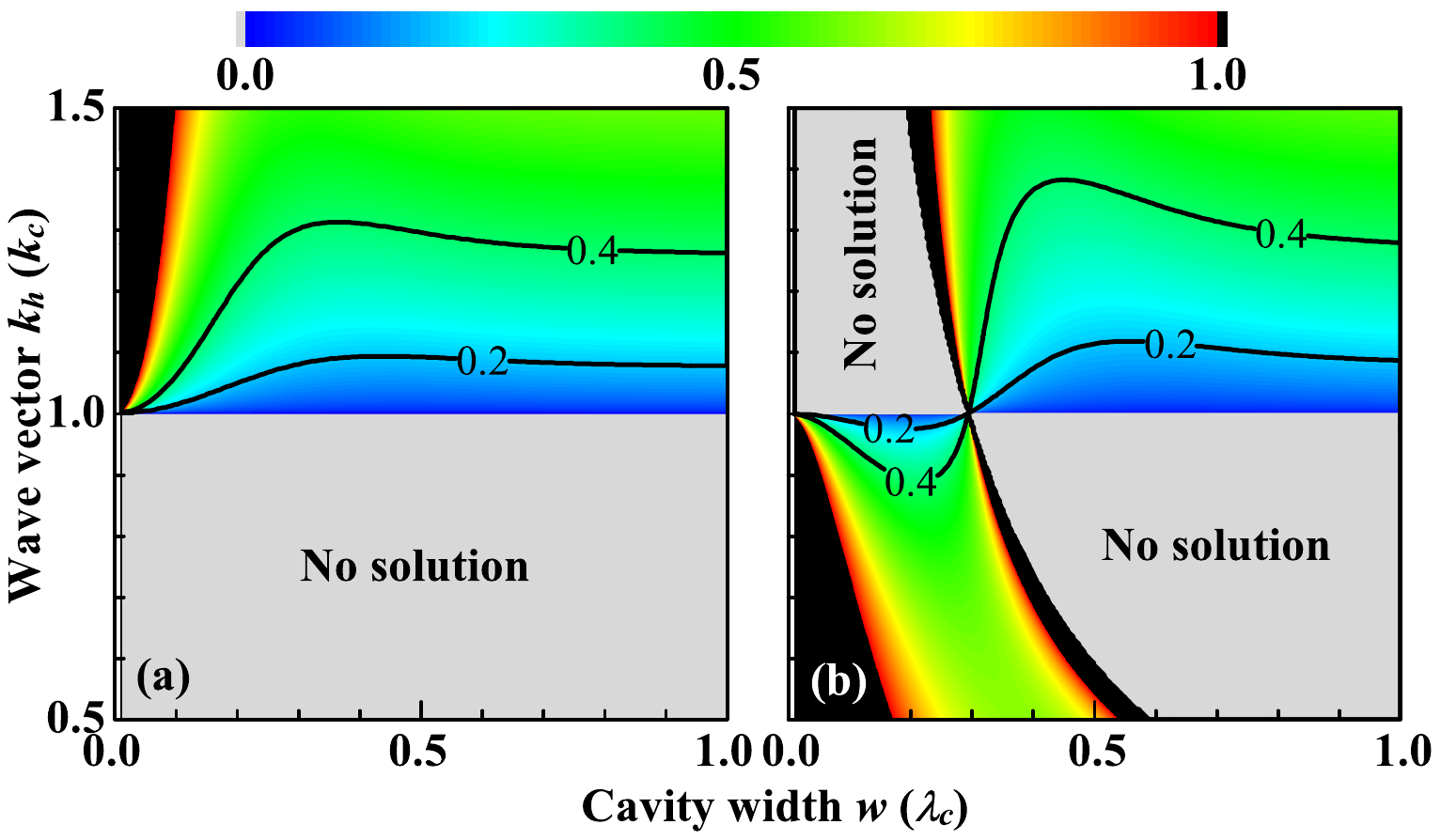}
  \caption{Contour plots of $V_0$ with the hybridization-state wave vector $k_h$ and the SOC width $w$ for the zeroth-order coupling (a) and the first-order coupling (b), respectively. In gray region, $V_0$ is negative which indicates that there does not exist the HS. The zeroth- and first-order coupling functions are defined in Eq.~(23). In addition, the unit of $V_0$ is $\omega_c$. In both panels, $\gamma_c=0.001\omega_c$.}
\end{figure}

Before discussing the HS, we first review the well-studied on-site coupling function, namely, $V(x)\propto\delta(x)$. As $V(x)\propto\delta(x)$, the transmission dip of the 1DW locates at the eigenfrequency of the SOC mode (in the present work, being $\omega_c$). It implies that the energy of the HS equals to the one of the SOC mode when $V(x)\propto\delta(x)$. Because $\delta(x)$ is an even function, the case of $V(x)\propto\delta(x)$ should be the limit of the nonlocal ZOC given in Eq.~(23a). This is confirmed by Eq.~(25a) from which one can easily find that $k_h=k_c$ when $w$ tends to zero. This also can be recognized from Fig.~2(a) which shows that $k_h$ indeed equals $k_c$ when $w$ tends to zero (no matter what value of $V_0$). The fact that $V_0^2$ is negative in the gray region of Fig.~2(a) indicates that there is no hybridization between the 1DW and SOC when $k_h<1.0$. This means that the HS wave vector $k_h$ is always larger than $k_c$ for the ZOC, and thus the HS is blue shifted compared with the SOC mode. The function $\Delta_k$ increases and then decreases with increasing the SOC width $w$ for the ZOC, referred to Eq.~(13c). As a result, the blueshift value of the HS energy also increases and then decreases with increasing $w$, referred to the contour lines in Fig.~2(a).

In Fig.~2(b), we also plot the contour graph of $V_0$ as the function of $V_0$ and $w$ for the FOC. Compared with Fig.~2(a), the HS for the FOC is much different from the ZOC. For the FOC, there exist two regions that do not support the HS, namely, the upper-left and lower-right corners of Fig.~2(b) where $V_0^2$ is negative. The touch point of these two regions locating at $k_h^f=k_c$ and $w\approx0.294\lambda_c$, represents a solution of Eq.~(25b) with uncertain $V_0$. That is to say, no mater what value of $V_0$ is, the value of $k_h^f=k_c$ always satisfies Eq.~(25b) as long as $w\approx0.294\lambda_c$. This is proved by the contour lines shown in Fig.~2(b): all the $V_0$ contour curves cross the point of $(k_c,~0.294\lambda_c)$. Therefore, we can roughly say that the HS energy is red and blue shifted for the FOC when $w$ is shorter and longer than $0.294\lambda_c$, respectively.

In all dark regions of Figs.~2(a) and 2(b), the coupling $V_0$ is larger than $\omega_c$. The condition of $V_0>\omega_c$ is hard to be achieved in practical cases, because the limitation of the SOC to the photon becomes weak when the coupling between the 1DW and SOC is too strong. Thus, we do not discuss them in the present work.

\begin{figure}
  \centering
  \includegraphics[width=8.5 cm]{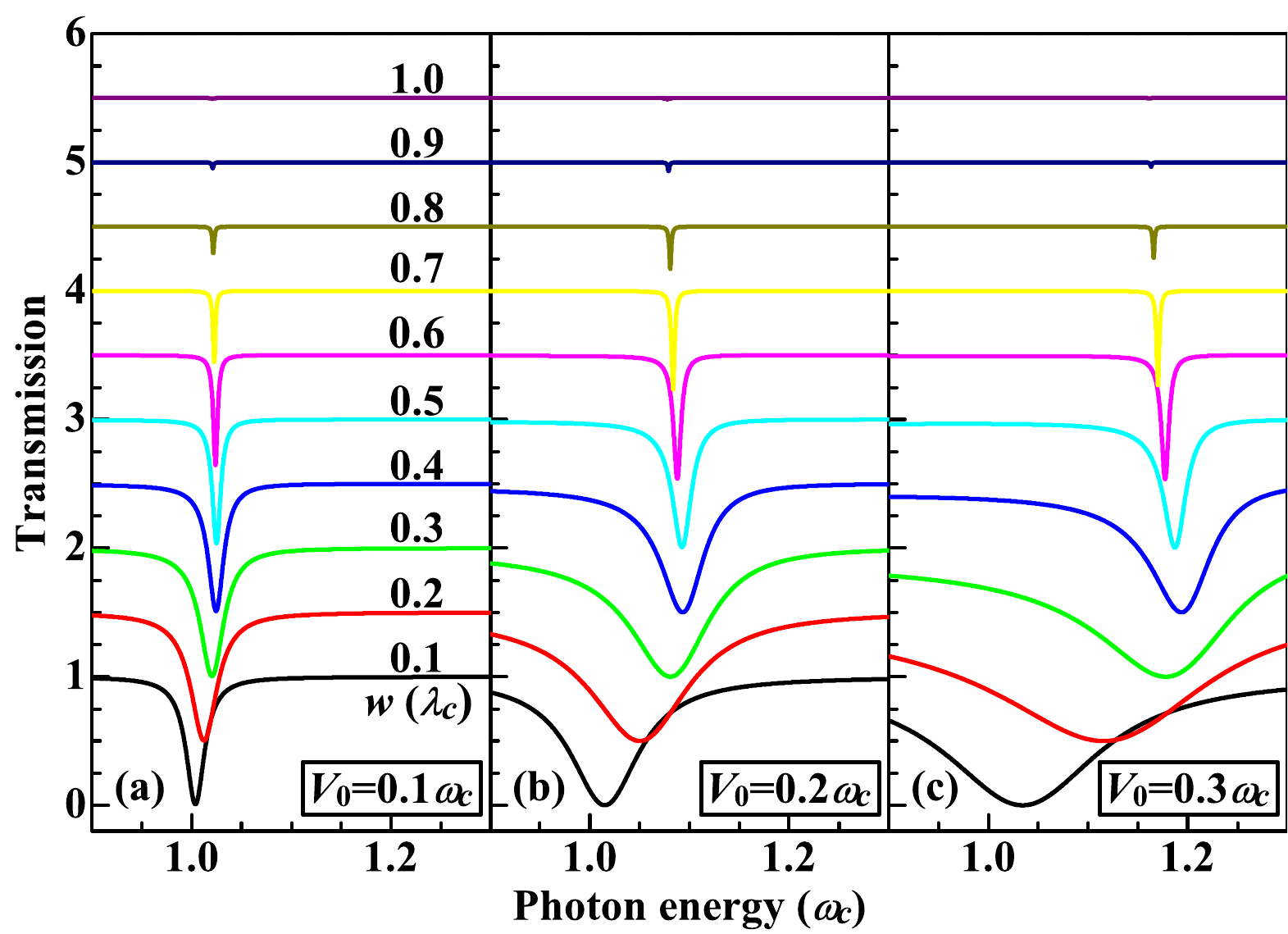}
  \caption{Transmission of the 1DW-SOC structure for the zeroth-order coupling. For easy observation, lines are offset from bottom with the step value being 0.5 in all panels. The energy and length units are $\omega_c$ and $\lambda_c=2\pi/k_c=2\pi v_g/\omega_c$, respectively. In all panels, $\gamma_c=0.001\omega_c$}
\end{figure}

In order to definitely show the influence of the nonlocal coupling to the single-photon transmission, we plot the single-photon transmission spectra in Fig.~3 for the ZOC. The curves in all panels are offsetted by a step being 0.5 for easy observation. The transmission dip position determined by the HS energy first shifts rightward and then leftward with increasing the SOC width $w$. This variation behavior of the transmission dip is consistent with the variation of the HS energy with the SOC width $w$ given in Fig.~2(a), and can be enhanced by increasing the coupling $V_0$, referred to Figs.~3(a-c). Simultaneously, the transmission dip is broadened, that is, the dip full width at half depth (FWHD) becomes large with increasing $V_0$. This is due to that $J_k$ (measuring the coupling between the 1DW and SOC modes) is proportional to $V_0^2$, referred to Eqs.~(21) and (24a).

From Fig.~3(c), one can further find that the transmission dip is not symmetric about the vertical axis. This is mainly determined by the dependence of $J_k$ on the wave vector $k$. Seen from Eqs.~(21) and (24a), $J_k$ decreases with increasing $k$ as well as the photon energy $\varepsilon$. Thus, the photon transmission on the left side of the transmission dip is lower than the right one, referred to Eq.~$(20)$. Naturally, this is from the nonlocal coupling between the 1DW and SOC. From Eqs.~(21) and (24a), we can find that the coupling $J_k$ takes the maximum value of $\sqrt{2\pi/e}V_0^2/(v_gk)$ when $w=\sqrt{2}k^{-1}$ for the ZOC. When $k$ is near $k_c$, the maximum $J_k$ is about at $w\approx0.23\lambda_c$. Hence, $J_k$ decreases with increasing $w$ when $w>0.23\lambda_c$. Consequently, the transmission dip becomes shallower and shallower with increasing the SOC width, as shown in Fig.~3. When $w$ tends to zero, the ZOC transits to the on-site coupling. In such a situation, the limit $\left.{J_k\over\Delta_k}\right|_{w\rightarrow0}={\sqrt{\pi}\over kw}$ which implies that there exists the coupling between the 1DW and SOC as well as the transmission dip even when $w$ tends to zero. It is also the reason why the strongly local coupling between the 1DW and SOC exists and can be described by the Dirac function.

\begin{figure}
  \centering
  \includegraphics[width=8.5 cm]{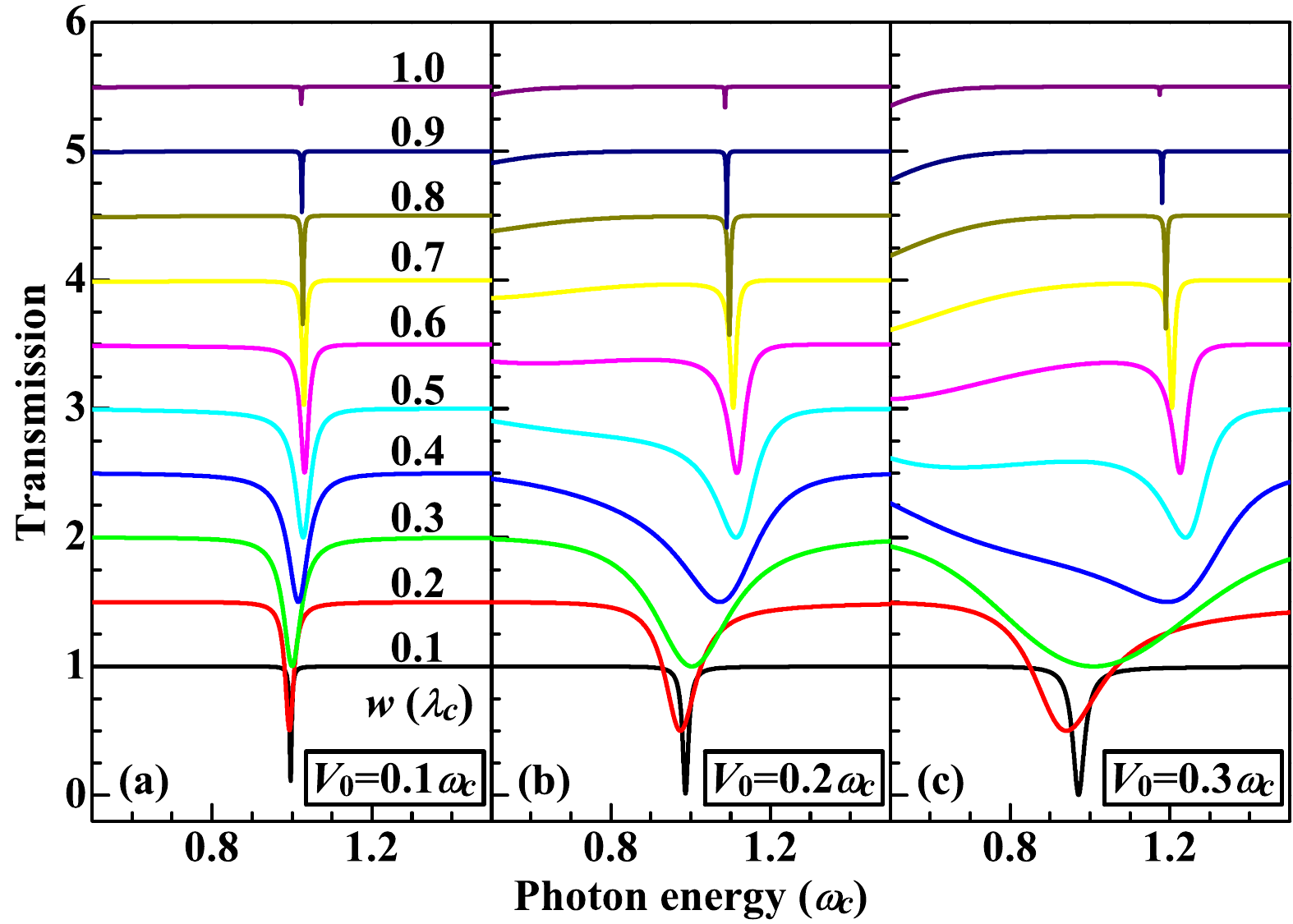}
  \caption{Transmission of the 1DW-SOC structure for the first-order coupling. For easy observation, lines are offset from bottom with the step value being 0.5 in all panels. The energy and length units are $\omega_c$ and $\lambda_c=2\pi/k_c=2\pi v_g/\omega_c$, respectively. In all panels, $\gamma_c=0.001\omega_c$.}
\end{figure}

The transmission spectra for the FOC are shown in Fig.~4. The shift of the transmission dip with the SOC width is consistent with the HS energy shown in Fig.~2(b): when $w<0.294\lambda_c$, they both red shift; oppositely, they both blue shift. For example, as $V_0=0.2\omega_c$, the HS energy obtained from Eq.~(25b) is $0.975\omega_c$ and $1.116\omega_c$ when $w=0.2\lambda_c$ and $0.6\lambda_c$, respectively [refer to Fig.~2(b)]. Correspondingly, the position of the transmission minimum value in Fig.~4(b) locates at $0.975\omega_c$ and $1.114\omega_c$ for the cases of $w=0.2\lambda_c$ and $0.6\lambda_c$. Increasing $V_0$ could enhance this kind of variation because $\Delta_k$ is proportional to $V_0^2$, as plotted in Figs.~4(a-c).

For $J_k\propto V_0^2$, the FWHD of the transmission dip also increases with increasing $V_0$ for the FOC. Similar with the ZOC, the depth of the transmission dip in Fig.~4 becomes shallower and shallower with increasing the SOC width when $w$ is large enough. This could be explained by a similar reason: $J_k$ for the FOC also first increases and then decreases with increasing $w$, and the transition value of $w$ equals to $\sqrt{6}k^{-1}$. Near the point of $k=k_c$, the transition value of $w$ is about $0.39\lambda_c$. This means that the transmission dip becomes shallower with increasing $w$ when $w\gtrsim0.39\lambda_c$. When $V_0=0.3\omega_c$, the transmission spectra in Fig.~4(c) display large difference from the ZOC, referred to Fig.~3(c). This is due to the more complexity of the dependence of $J_k$ and $\Delta_k$ on $k$ for the FOC compared with the ZOC. In the discussion on the ZOC, we have stated that the coupling between the 1DW and SOC exists even at the limit of $w\rightarrow0$. For the FOC, it is fully different. When $w\rightarrow0$, one can find $\left.{J_k\over\Delta_k}\right|_{w\rightarrow0}=-{\sqrt{\pi}kw\over2}$ for the FOC. Here the minus sign dates from the fact that the HS is red shift compared with the SOC mode when $w$ is small, referred to Fig.~2(b). Obviously, when $w$ tends to zero, the limit $\left.{J_k\over\Delta_k}\right|_{w\rightarrow0}$ equals zero, which indicates that there does not exists the coupling between the 1DW and SOC for the FOC when the SOC is too narrow.

\begin{figure}
  \centering
  \includegraphics[width=8.0 cm]{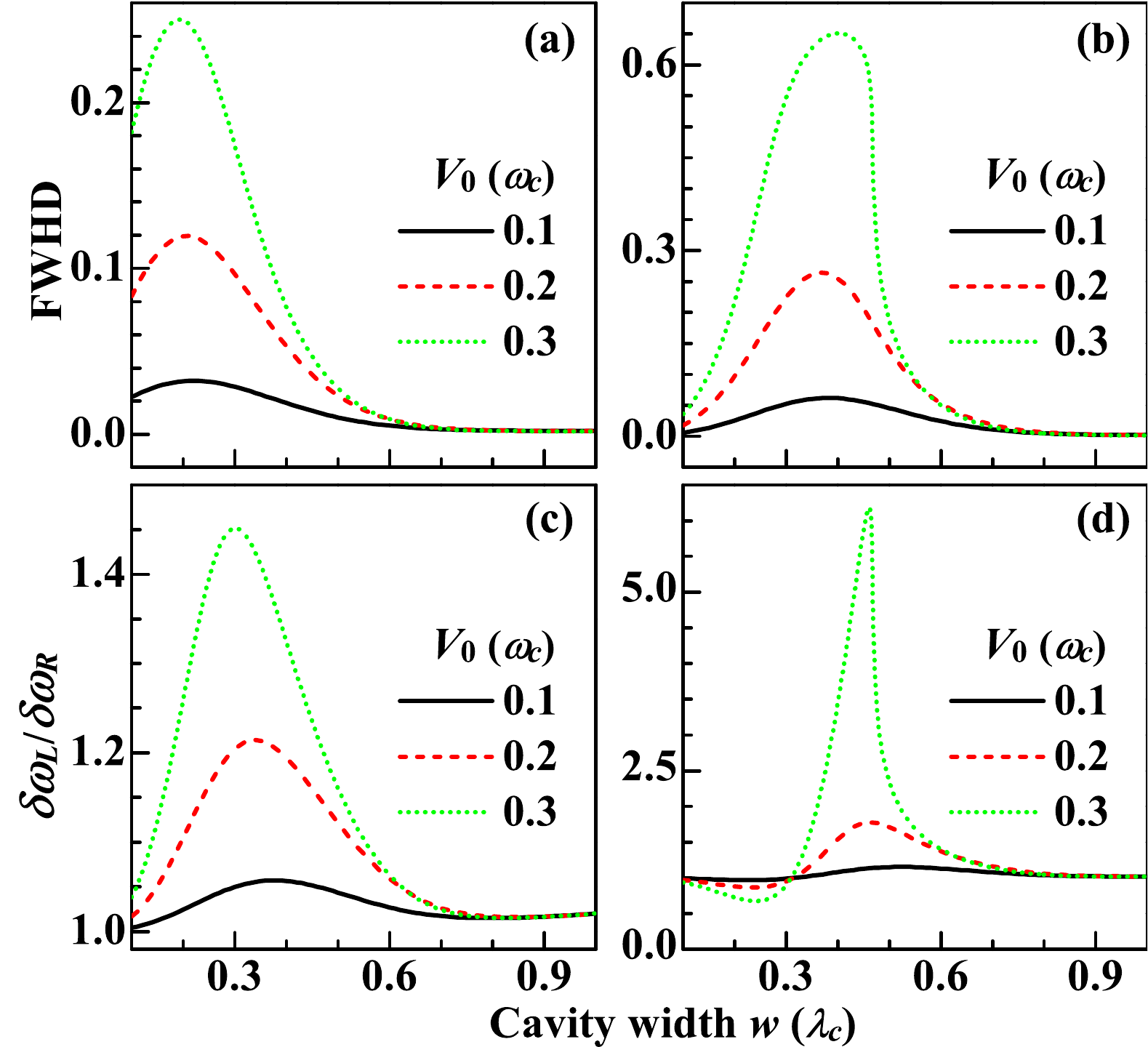}
  \caption{Variation of the full width at half depth of the transmission dip with the SOC width for the zeroth-order coupling (a) and for the first-order coupling (b). In (c) and (d), the ratios of the dip left half width at half depth, $\delta\omega_L$, to the right one,  $\delta\omega_R$, are plotted for the zeroth- and first-order couplings, respectively. In all panels, $\gamma_c=0.001\omega_c$.}
\end{figure}

In order to study the spectra shape in detail, we plot the dip full width at half depth (FWHD) in Figs.~5(a) and 5(b) for the ZOC and FOC, respectively. From them, we can definitely find that the FWHD first increases and then decreases with increasing the SOC width. Simultaneously, this behavior of the FWHD could be enhanced by increasing $V_0$. They are, respectively, due to the facts that $J_k$ holds maximum for $w$ and that $J_k$ is proportional to $V_0^2$ for both coupling functions.

For describing the asymmetry of the transmission dip, we introduce two parameters: the dip left half width at half depth, $\delta\omega_L$, and the dip right half width at half depth, $\delta\omega_R$. For the symmetric transmission dip, they are equal to each other. The ratios of $\delta\omega_L$ to $\delta\omega_R$ are plotted in Figs.~5(c) and 5(d) for the ZOC and FOC, respectively. For the ZOC, $\delta\omega_L$ is always larger than $\delta\omega_R$, because $J_k$ always decreases with increasing $k$ no matter what value of $w$. It is different from the FOC whose $J_k$ increases with increasing $k$ when $w$ is small. Thus, $\delta\omega_L$ is smaller than $\delta\omega_R$ for the FOC when $w$ is small, as shown in Fig.~5(d). When $w$ is large, it is the same with the ZOC that $J_k$ decreases with increasing $k$. In addition, the dip asymmetry is much more apparent for the FOC than for the ZOC, which should be attributed to that the FOC holds stronger delocalization than the ZOC. Increasing $V_0$ also enhances the transmission-dip asymmetry. Compared with the symmetric transmission spectra for the on-site coupling, the asymmetry of the transmission spectra here is fully from the nonlocal coupling between the 1DW and SOC.

\begin{figure}
  \centering
  \includegraphics[width=8.5 cm]{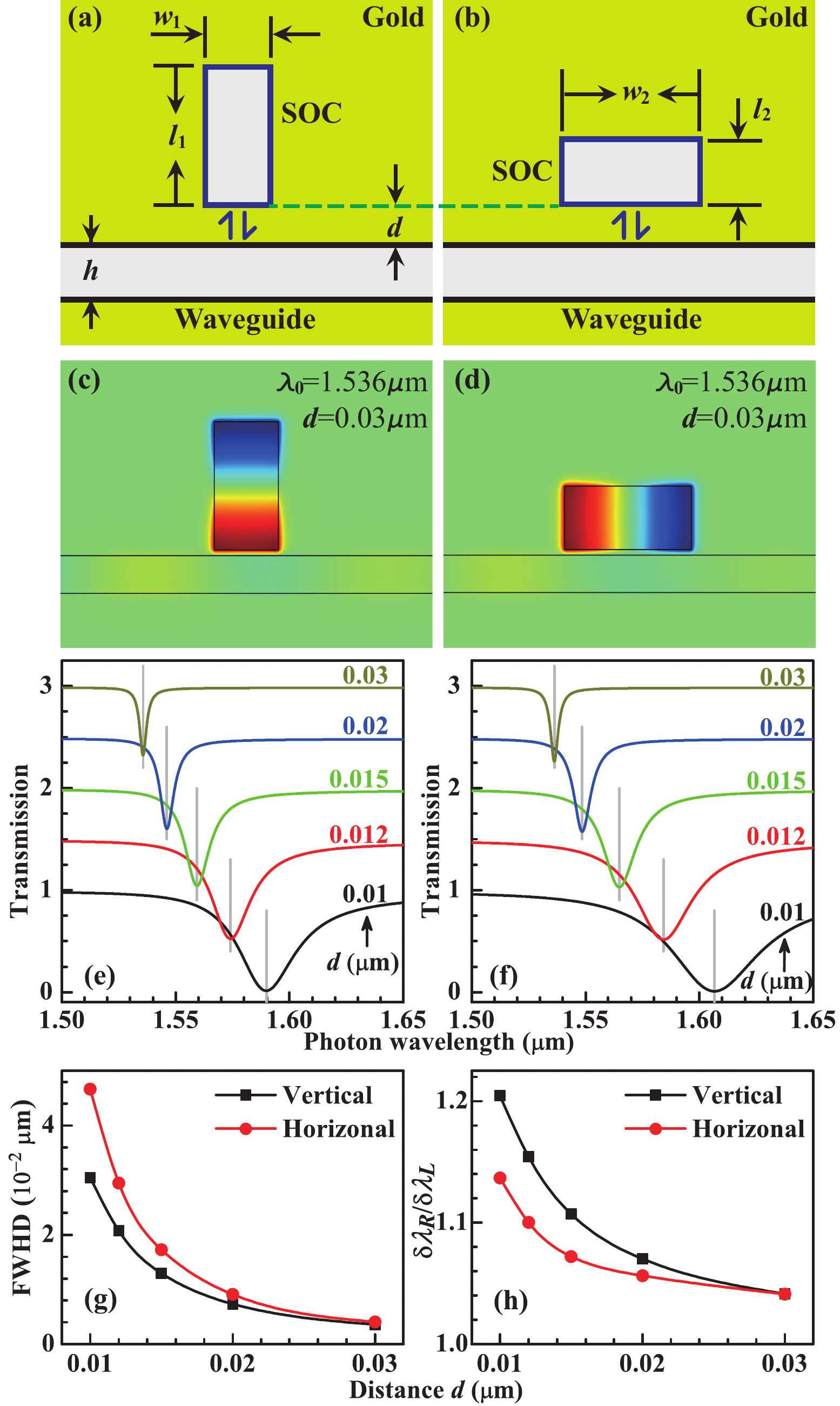}
  \caption{Drafts of the gold-based 1DW-SOC are shown in (a) and (b) where the SOCs are arranged vertically and horizonally, respectively. (c) and (d) show the magnetic distribution of the magnetic field when the vacuum wavelength of the incident photon $\lambda_0=1.536~\mu$m for the vertical and horizonal situations, respectively. The photon transmission is shown in (e) and (f) with different distance $d$ between the 1DW and SOC. For clarity, the transmission lines are offseted with the step being 0.5. Simultaneously, the positions of the transmission dips are denoted by the short gray lines. In (g) and (h), we plot the dip full width at half depth and the ratios of the dip right half width at half depth, $\delta\lambda_R$, to the left one,  $\delta\lambda_L$, respectively. Please note that here they are defined through the wavelength instead of the frequency. The structure parameters denoted in (a) and (b) are: $h=0.2~\mu$m, $w_1=l_2=0.34~\mu$m, and $l_1=w_2=0.68~\mu$m.}
\end{figure}

\section{Simulated by the gold-based 1DW-SOC}

In the present section, we take the gold-based 1DW-SOC structure to simulate the proposed nonlocal effects. The considered structure of the 1DW-SOC are shown in Figs.~6(a) and 6(b) where the structure parameters are denoted and given in the caption. The SOC mode along the short-side direction is an even mode, while the SOC mode along the long-side direction is an odd mode. This can be clearly seen from Figs.~6(c) and 6(d) which show the magnetic field distribution of the the electromagnetic wave with the vacuum wavelength $\lambda_0$ being 1.536 $\mu$m. Consequently, the coupling between the 1DW and SOC can be approximated by the ZOC and FOC for the structures shown in Figs.~6(a) and 6(b), respectively. From the magnetic field distribution, one can read the wavelength of the 1DW mode $\lambda$ is about 1.3 $\mu$m. Thus, $w/\lambda\approx0.26$ and $l/\lambda\approx0.52$ for the vertical and horizonal arrangements, respectively. They imply that the nonlocal coupling effects could be observed for both arrangements. In addition, we should point out that the above approximation is not rather strict, though it can be done according to the distribution of the SOC mode. This is due to that the magnetic field in the cavity is almost uniform along the short-side direction and is more nonlocal along the long-side direction than the FOC. Hence, the transversal effective sizes are a little longer than $w_1$ and $w_2$ for the vertical and horizonal arrangements, respectively.

In Figs.~6(e) and 6(f), the transmission spectra are plotted under several distances between the 1DW and SOC. The interaction between the 1DW and SOC decreases with increasing $d$. Both figures 6(e) and 6(f) show that the transmission dip shifts rightward (redshift) with decreasing $d$. Such phenomenon is not consistent with our discussion on the ZOC and FOC examples shown in last section. This conflict is due to that the SOC eigenfrequency also red shifts with decreasing $d$: the limitation of the SOC on the photon becomes weaker and weaker with decreasing $d$. It can strongly shield the blue shift of the transmission dip due to the nonlocal coupling. However, the shift of the SOC eigenfrequency does not strongly influence the line shape, including the FWHD and asymmetry of the transmission dip. They are plotted in Figs.~6(g) and 6(h), respectively. Please note that in Figs.~6(g) and 6(h) the FWHD and right or left half width at half depth are defined through the wavelength instead of the frequency.

Figure 6(g) shows that the FWHD decreases with increasing $d$ for both the vertical and horizonal arrangements which are consistent with the ZOC and FOC situations. The ratio of the FWHD to the photon wavelength of the SOC mode is smaller than 3\% for all points in Fig.~6(g). Together with the parameter relations of $w_1/\lambda\approx0.26$ and $w_2/\lambda\approx0.52$, from Figs.~5(a) and 5(b) we could estimate that the coupling between the 1DW and SOC is about one tenth of the SOC eigenfrequency when $d=0.012~\mu$m for the structures shown in Figs.~6(a-b). From Figs.~5(a) and 5(b), we can further find that the ratio of the FOC's FWHD with $w\approx0.52\lambda_c$ to the ZOC's with $w\approx0.26\lambda_c$ is about 1.3 when $V_0=0.1\omega_c$. This ratio also can be obtained from Fig.~6(g), and is about 1.4 when $d=0.012~\mu$m. They are qualitatively consistent with each other. In addition, the ratios of $\delta\lambda_R/\delta\lambda_L$ plotted in Fig.~6(h) are larger than 1 for both vertical and horizonal arrangements, which are also qualitatively consistent with the results of the ZOC and FOC given in Figs.~5(c) and 5(d). These simulation results confirm that the nonlocal coupling effects can really appear in the coupled structure of the 1DW and SOC.

\section{conclusion}

In the present work, we studied the effects of the nonlocal coupling between the one-dimensional waveguide and the side optical cavity. The real-space Hamiltonian with nonlocal coupling terms was derived in the quantum field view. From the Hamiltonian, we found an analytical formula for the single-photon transmission. Simultaneously, an equation was derived to determine the nontrivial state --- the hybridization state between the one-dimensional waveguide mode and the side optical cavity mode. Two examples of nonlocal coupling, namely, the zeroth- and first-order couplings, were adopted to show the effects due to the nonlocal coupling. From them, we got that the hybridization state is blue shift for the zeroth-order coupling compared with the side cavity mode, while the hybridization mode is red shift when the transversal cavity scale is short. The nonlocal coupling also induce the asymmetry of the single-photon transmission dip. At last, we simulated this nonlocal coupling effects by the gold-based one-dimensional waveguide. Their transmission behaviors are consistent with the conclusions of the zeroth- and first-order couplings. These results could provide a way to design practical one-dimensional waveguide devices through controlling their interactions with the side optical cavities.

\section*{Acknowledgements}
This work is supported by NSFC (Grant No. 11304015), Beijing Higher Education Young Elite Teacher Project (Grant No. YETP1228), BIT Foundation for Basic Research (Grant No. 20100942018).

\bibliographystyle{aipnum4-1}
%

\end{document}